\documentclass[12pt]{article}

\usepackage{scicite}

\usepackage{graphicx}
\usepackage{epsfig,amsmath,amssymb,bm,epsf,psfrag,bbm}
\usepackage{graphicx}
\usepackage{dcolumn}
\usepackage{bm}
\usepackage{amsbsy}
\usepackage{amsthm}
\usepackage{color, comment, verbatim}
\usepackage{ragged2e}
\usepackage{subfigure}
\usepackage{enumerate}
\usepackage{chngcntr}
\usepackage[T1]{fontenc}
\usepackage[english]{babel}
\usepackage[utf8]{inputenc}
\usepackage[sectionbib]{chapterbib}
\usepackage{lineno}

\usepackage{dsfont} 

\usepackage[linesnumbered,lined,ruled,commentsnumbered]{algorithm2e}

\usepackage{tikz}
\usepackage[toc,page]{appendix}
\usepackage{titlesec}
\usepackage{booktabs}
\usepackage{tablefootnote}
\usepackage{threeparttable}
\usepackage{tikz}
\usepackage[colorlinks,breaklinks]{hyperref} 
\usepackage{xcolor}

\usepackage{times}

\newenvironment{sciabstract}{%
\begin{quote} \bf}
{\end{quote}}

\newcounter{lastnote}

\title{Single-pixel transmission matrix recovery via 2-photon fluorescence} 

\author
{Shupeng Zhao,$^{1, 2}$ Bernhard Rauer,$^{1}$ Lorenzo Valzania,$^{1}$ Jonathan Dong,$^{3}$ \\ 
Ruifeng Liu,$^{2}$ Fuli Li,$^{2}$ Sylvain Gigan,$^{1}$ Hilton B. de Aguiar,$^{\ast1}$\\
\\
\normalsize{$^{1}$Laboratoire Kastler Brossel, ENS-Université PSL, CNRS, Sorbonne Université,}\\
\normalsize{Collège de France. 24 rue Lhomond, 75005 Paris, France.}\\
\normalsize{$^{2}$Shaanxi Province Key Laboratory for Quantum Information and Quantum Optoelectronic Devices,}\\
\normalsize{and Department of Applied Physics, School of Science, Xi’an Jiaotong University,}\\
\normalsize{Xi’an 710049, China}\\
\normalsize{$^{3}$Biomedical Imaging Group, Ecole polytechnique fédérale de Lausanne (EPFL).}
\\
\normalsize{$^\ast$E-mail:  h.aguiar@lkb.ens.fr.}
}

\date{}

\begin{document}

\baselineskip24pt

\maketitle

\begin{sciabstract}
Imaging at depth in opaque materials has long been a challenge. Recently, wavefront shaping has enabled significant advance for deep imaging. Nevertheless, most non-invasive wavefront shaping methods require cameras, lack the sensitivity for deep imaging under weak optical signals, or can only focus on a single "guidestar". Here, we retrieve the transmission matrix (TM) non-invasively using two-photon fluorescence exploiting a general single-pixel detection framework, allowing to achieve single-target focus on multiple guidestars spread beyond the memory effect range. In addition, if we assume memory effect correlations exist in the transmission matrix, we are able to significantly reduce the number of measurements needed. 
\end{sciabstract}

\section*{Introduction}

Fluorescence-based microscopy is the workhorse imaging modality in biology and medicine. However, imaging opaque biological specimens at-depth is a current challenge in optics \cite{bertolotti2022imaging,gigan2022roadmap} as light passing through biological tissue experiences an environment of heterogeneous refractive index, leading to aberrations and scattering. To overcome these effects, feedback-based wavefront shaping techniques have emerged as a new paradigm to inverse the effect of scattering and focus light to a diffraction-limited spot \cite{vellekoop2007focusing}. This is achieved by iteratively modulating the input wavefront with a spatial light modulator (SLM) aided by a feedback signal from the targeted point. However, this correction is only valid for one focus position and the process has to be repeated for another position. Retrieving the transmission matrix (TM), which connects input and output planes of the scattering medium, brings more freedom to focus light at any position of the target plane \cite{popoff2010measuring}.
Despite significant advances in imaging by characterizing the transmission matrix of a system, there are many applications scenarios where this cannot be neatly done by simply imaging the output target plane. Non-invasive settings required in biological imaging, typically using epi-geometry, are more challenging as they require an undisturbed feedback from within the medium for optimization- or TM-based schemes. In such cases, wavefront optimization requires the assistance of beacons or so-called "guidestars" in the target plane \cite{horstmeyer2015guidestar}. Photoacoustic feedback \cite{kong2011photoacoustic,chaigne2014controlling}, and acousto-optic tagging \cite{tay2014ultrasonically,si2012fluorescence,ruan2014iterative,wang2012deep,tang2012superpenetration,xu2011time} are compelling methods because of acoustic waves's resilience to scattering \cite{xu2011time}, which enables the reconstruction of a TM but requires complex super-resolution methods to enhance the resolution that is limited by the sound wavelength \cite{resolution_katz2019controlling,resolution_conkey2015super,resolution_judkewitz2013speckle,resolution_si2012breaking}. Another powerful method relies on the measurement of a time-gated matrix in reflection \cite{badon2016smart,jeong2018focusing}, but is based on retro-reflected ballistic excitation photons, hence limited in depth because of their exponential attenuation.

Fluorescence-based feedback signals can be easily separated from the excitation wavelength, however reach focusing at the optical diffraction limit has been a major challenge \cite{horstmeyer2015guidestar}. Because the fluorescence signal is strongly scattered by the complex medium, the main difficulty in wavefront shaping experiments is how to assess whether the excitation energy was delivered to only one guidestar. Optical non-linear fluorescence \cite{katz2014noninvasive}, or other computational non-linear methods~\cite{boniface2019noninvasive,aizik2022fluorescent,aizik2023non} have been proposed to solve this problem. Once achieved focusing on a single guidestar, memory effect (ME) of the scattering medium \cite{feng1988correlations,freund1988memory} can be used for raster-scanning imaging \cite{katz2014noninvasive}. Nevertheless, the imaging range is limited by the ME range. A potential solution to overcome this limitation is to recover the TM using camera-based methods, which have been used to demix the emission signal from the different guidestars \cite{boniface2020non,d2022physics}. Summarizing current status~\cite{gigan2022roadmap}, previous fluorescence based strategies can either focus at one single position using single-pixel detectors \cite{tang2012superpenetration,liu2015optical,katz2014noninvasive,papadopoulos2017scattering,papadopoulos2020dynamic,may2021fast,daniel2019light} (or cameras \cite{aizik2022fluorescent,boniface2019noninvasive}), or multiple positions by reconstructing the TM using cameras \cite{d2022physics,boniface2020non,baek2023generalized}.Thus, a method based on single-pixel detection that can reconstruct the TM is urgently needed, as camera-based methods lack sensitivity at-depth (due to weaker signal generation).

Here, we propose a general model of fluorescence microscopy to perform non-invasive focusing and imaging at multiple target positions through the retrieval of the coherent TM, using only a single-pixel detector. We start by establishing a general framework and show that the forward model is only solvable with non-linear fluorescence signals. In particular, we use a multiplexed phase retrieval scheme using gradient-descent-based algorithm to extract the TM. We demonstrate focusing on multiple fluorescing targets beyond the memory effect. To speed up the measurement process, we consider the memory effect which significantly decreases the amount of measurements needed.  

\section*{Principle}

\begin{figure*}[htp]
	\centering
	\includegraphics[width=1\columnwidth]{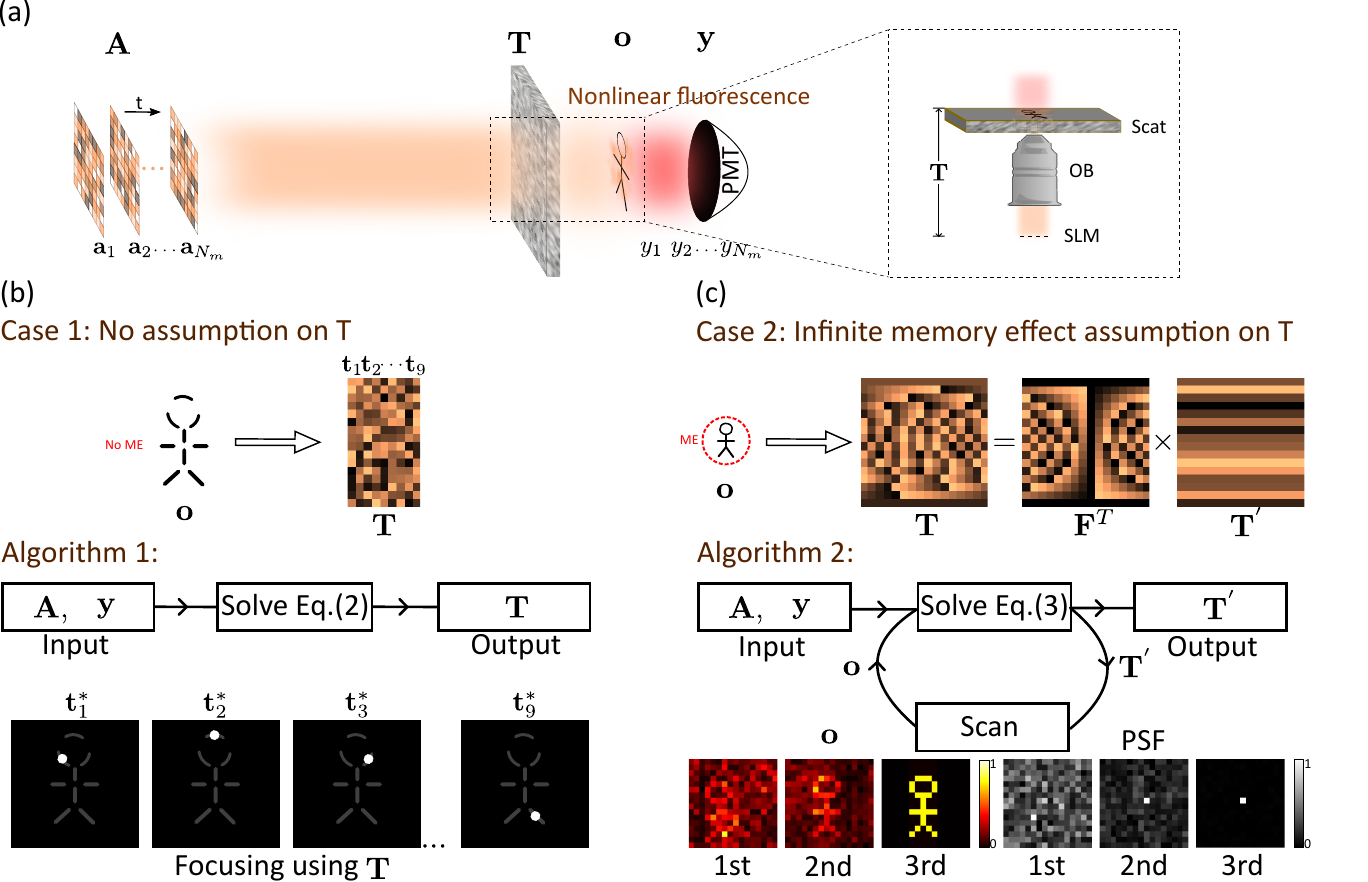}
	\caption{\textbf{Experimental and reconstruction principles.} (a) Simplified schematic view of a general fluorescence microscope with a single-pixel detector. Random wavefront generated by a SLM (each wavefront represents one row of matrix $\mathbf{A}$) impinges on a scattering medium with a transmission matrix $\mathbf{T}$ and excites 2-photon fluorescence (2PF) of the extended object $\mathbf{o}$. The fluorescence signal $\mathbf{y}$ is collected by a single-pixel detector (photo-multiplier tube (PMT)). Retrieval of the TM is achieved in two different extreme cases 1 and 2, where tailored algorithms are used for solving different models. (b) In case 1, $\mathbf{T}$ is assumed to have completely uncorrelated elements, i.e. a random matrix. We solve for $\mathbf{T}$ using as input the measurements $\mathbf{y}$ and the wavefront modulation embedded in the user-input matrix $\mathbf{A}$. The reconstruction of $\mathbf{T}$ allows to focus on each guidestar in the object. (c) In case 2, $\mathbf{T}$ has correlations, i.e. the infinite ME range is considered, and is given by the pixel-wise product of a rank-1 matrix $\mathbf{T}^{'}$ and a "tilt matrix" $\mathbf{F}^{T}$, where $\mathbf{F}$ is a matrix corresponding to 2D Fourier transform. An iterative method is used where, in each iteration, we solve for $\mathbf{T}^{'}$ at first, then the wavefronts estimation are used for raster-scanning the sample to update $\mathbf{o}$ until convergence.
 }
	\label{fig:fig1}
\end{figure*}

We are interested in the situation described in Fig \ref{fig:fig1}a: an extended incoherent object hidden behind a scattering medium is illuminated by a series of random speckle patterns, created by modulating an incident laser via spatial light modulator. The total fluorescence is integrated by a single-pixel detector. That is, we do not resolve the spatial fluctuations of the fluorescence pattern. We can model this in a general matrix-form phase-retrieval framework, where the detected signal $\mathbf{y} \in \mathbb{R}^{N_{\rm{m}}\times1} $ is described by
\begin{equation}
\mathbf{y} = \lvert \mathbf{A}\mathbf{T} \rvert^{2p} \mathbf{o},
\label{eq:eq1}
\end{equation}
where $\mathbf{A} \in \mathbb{C}^{N_{\rm{m}}\times N_{\rm{slm}}}$ is a known measurement matrix, with each of the $N_{\rm{m}}$ row representing a realization of modulated wavefront defined by $N_{\rm{slm}}$ pixels, $\mathbf{T} \in \mathbb{C} ^{N_{\rm{slm}}\times N_{\rm{obj}}} $ is a field transmission matrix from the SLM to the fluorescent object plane, $\mathbf{o} \in \mathbb{R}^{N_{\rm{obj}}\times1} $, and $p$ the order of the $p$-photon excited fluorescence signal. In what follows, we consider two extreme cases: thick scattering medium with negligible ME range (the columns in $\mathbf{T}$ are fully decorrelated, case 1 in Fig.\ref{fig:fig1}b) and thin scattering medium with infinite ME range (the columns in $\mathbf{T}$ are fully correlated, case 2 in Fig.\ref{fig:fig1}c).

The object being sparse, only the positions where the fluorescent response of the object is non-zero contribute to the signal. Therefore we can consider only some columns of $\mathbf{T}$ where the fluorescent response is non-zero. To do that, we merge the vector $\mathbf{o}$ into the matrix $\mathbf{T}$ and Eq. \eqref{eq:eq1} becomes:
\begin{equation}
\mathbf{y} = \sum_{i=1}^{N^{'}_{\rm{obj}}} \lvert \mathbf{A} \mathbf{t}_{i} \rvert^{2p},
\label{eq:eq2}
\end{equation}
where $N^{'}_{\rm{obj}} < N_{\rm{obj}}$ represents the number of columns left after merged with $\mathbf{o}$. $\mathbf{t}_{i}$ represent the columns of the transmission matrix $\mathbf{T}$ weighted by the corresponding elements in $\mathbf{o}^{\frac{1}{2p}}$. Note that $p=1$ corresponds to linear fluorescence, and in this case the solution of Eq.\eqref{eq:eq2} is not unique \cite{dong2019spectral}, because there are infinite matrices $\mathbf{T}$ that satisfy the same measurement $\mathbf{y}$ (see the Supplementary Information for detail). Conversely, for $p=2$ the ambiguity is lifted and Eq.\eqref{eq:eq2} has unique solution provided that a certain number of measurements is fulfilled (see the Supplementary Information for a mathematical proof). Therefore, as long as we detect nonlinear fluorescence signals and collect enough random wavefront measurements, the transmission matrix $\mathbf{T}$ can be uniquely determined. Note that $\mathbf{A}$ is, in our case, obtained by displaying a series of random patterns on a SLM, but could be achieved by any other deterministic modulation scheme (galvanometric mirrors, digital micromirror devices etc). 

In order to reconstruct $\mathbf{T}$ with 2-photon fluorescence signal, we use a gradient-descent-based algorithm to solve Eq.\eqref{eq:eq2} when $p=2$ (see methods for further details on the Algorithm 1). Since we are solving a multiplexed phase retrieval problem, the number of wavefronts measured $N_{\rm{m}}$ should scale with the number of controlled modes $N_{\rm{slm}}$ (the dimension of unknown signal) and the number of excitable speckles grains at the object plane where the fluorescence response is nonzero: $N^{'}_{obj}$ (the number of unknown signals). It means the required measurement number depends on the complexity of the object. 

While this method allows to find a solution, including the presence of memory effect into our forward model allows us to reduce the measurement number. As showing in Fig.\ref{fig:fig1}c, under the infinite memory effect condition, there is only phase ramp difference between columns of $\mathbf{T}$. These phase ramp can be modeled by a "tilt matrix" $\mathbf{F}^{T}$, where $\mathbf{F}$ is a matrix corresponding to 2D Fourier transform (Multiplying $\mathbf{F}$ with a vector is equivalent to first reshaping the vector into a matrix, then doing 2D Fourier transform, and finally reshaping it back into a vector.). We can then rewrite Eq.\eqref{eq:eq1} to
\begin{equation}
\mathbf{y} = \lvert \mathbf{A}(\mathbf{F}^{T}\odot \mathbf{T}^{'}) \rvert^{4} \mathbf{o},
\label{eq:eq3}
\end{equation}
where $\mathbf{T}^{'}$ is a matrix that every column is the same vector $\mathbf{t}$ that corresponds to a wavefront ($\odot$ stands for element-wise product). In order to retrieve both $\mathbf{T}^{'}$ and $\mathbf{o}$ in Eq.\ref{eq:eq3}, we propose a synergistic algorithm to ensure convergence. As schematically shown in Fig. \ref{fig:fig1}c: (1) collect measurement vector $\mathbf{y}$ with the measurement matrix $\mathbf{A}$, and initialize $\mathbf{o}$ by raster-scanning the sample, (2) update $\mathbf{t}$ by solving Eq.\eqref{eq:eq3} numerically, (3) update $\mathbf{o}$ by raster-scanning the sample with the compensate wavefront $\rm{arg}(\mathbf{t}^{*}) $, (4) repeat (2) and (3) until a strong 2-photon signal is detected. (see methods for further details on the Algorithm 2). Since the matrix $\mathbf{T}^{'}$ has only one vector $\mathbf{t}$ as unknown, the multiplexed phase retrieval problem simplifies to a normal phase retrieval problem with only one column to be reconstructed. Hence, the measurement number decreases considerably going from $N_{\rm{m}} \propto N_{\rm{slm}} \times N^{'}_{\rm{obj}}$ to typically $N_{\rm{m}} \propto N_{\rm{slm}}$. Note here that instead of only using an SLM to generate a measurement matrix $\mathbf{A}$, we can use a scan element and SLM together to generate $\mathbf{A}$: this allows to speed up measurements significantly (e.g. using galvanometric mirrors).

\section*{Results}

\begin{figure*}[ht]
        \centering
	\includegraphics[width=0.7\columnwidth]{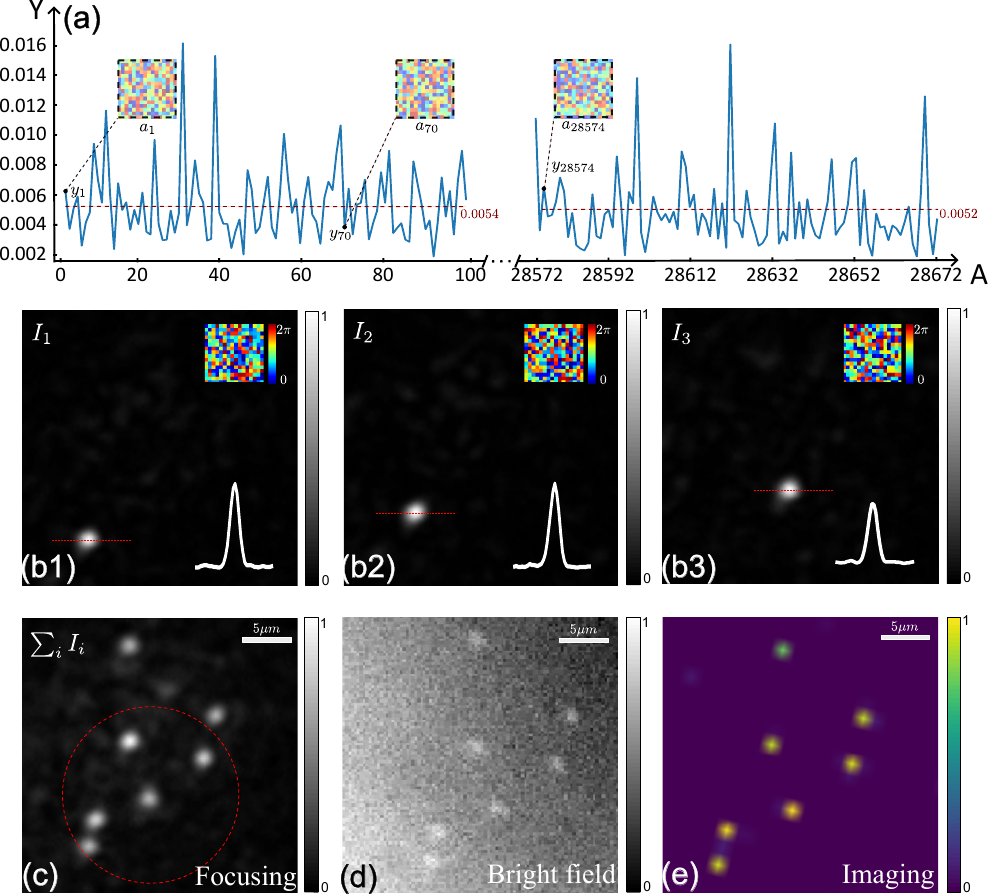}
	\caption{\textbf{$\mathbf{T}$ retrieval beyond the ME.} (a) 2PF signal detected ($\mathbf{y}$) vs. number of random wavefronts. Representative wavefronts used are shown in the insets. The two red dashed line represents the average level of the first and last 100 signal samples, respectively.  (b) Inspected focii (control CCD) for each column after retrieval of $\mathbf{T}$. The insets show line profile. (c) Sum of all images for each focus demonstrating unique single focus for each column retrieved. (d) Brightfield image taken with control camera confirms the number of sources. (e) Imaging after analysis on reconstructed $\mathbf{T}$ using correlations between the different columns (The red circle in (c) depicts the memory effect range).}
	\label{fig:fig2}
\end{figure*} 

To demonstrate the feasibility of the methods, we have performed initial experiments on sparse fluorescence samples. We modulate the wavefront sequence using a SLM to define measurement matrix $\mathbf{A}$  (see method for details on the setup).  The SLM is located conjugate to the back focal plane of the illumination objective. The fluorescence samples are drop-cast on a glass coverslip whose bottom surface, facing the illumination objective, was sand-blasted to introduce a single strongly scattering layer \cite{rauer2022scattering}. After the scattering medium and the sample, the 2PF signal $\mathbf{y}$ is collected by a photomultiplier tube (PMT). An additional charge-coupled device (CCD) camera is used only to inspect the quality of the focusing performance. A typical realization of the detected 2PF signal is shown in Fig. \ref{fig:fig2}a where the fluctuations are due to the random wavefront realizations: on short-term averages these fluctuations are constant not showing any bleaching, and are furthermore well contrasted. We then solve Eq.\eqref{eq:eq2} when $p=2$, from which we retrieve $\mathbf{T}$ to find the wavefronts $\rm{arg}(\mathbf{t}^{*}_{i})$ to focus on individual multiple targets (Fig.\ref{fig:fig2}b), or focii summed (Fig. \ref{fig:fig2}c) as inspected by the CCD camera. Bright field images of the sources match well with the results (Fig. \ref{fig:fig2}d). Residual correlations in $\mathbf{T}$ allows for imaging without raster-scanning the focus (Fig. \ref{fig:fig1}d) and beyond the memory effect range (the red dashed circle in Fig.\ref{fig:fig2}c represents the memory effect range). For this imaging example, we calculate the correlation of each two columns of $\mathbf{T}$. Computing successively these pairwise correlations between close targets allows retrieving the full object, well beyond a single ME range (see Supplementary Information for further details). Importantly, we did not use any assumption on the ME to retrieve the transmission matrix, but only to retrieve a posteriori the relative position between sources. 
 
\begin{figure}[htp]
    \centering
	\includegraphics[width=0.7\columnwidth]{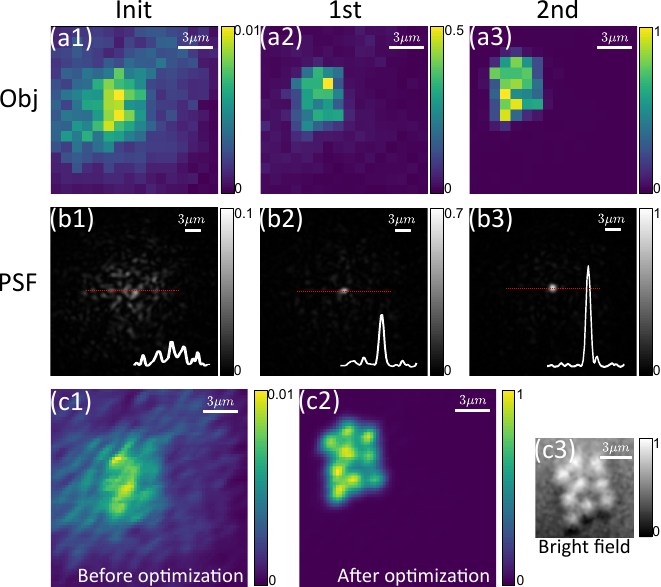}
	\caption{\textbf{$\mathbf{T}$ retrieval assuming ME correlations.} (a) 2PF imaging by raster scanning the sample using the speckles depicted in (b), themselves captured with the CCD before each iteration. The images are low-resolution in order to speed up acquisition. High resolution images before (c1) and after two iterations (c2) show the cluster, and its morphology confirmed by the brightfield camera (c3).}
	\label{fig:result8}
\end{figure} 

To speed up measurement and reconstruction time, we then studied a system that has ME in $\mathbf{T}$, and explicitly take it into account in the algorithm. Fig.\ref{fig:result8} shows the experimental result when 8 random patterns are used for raster-scanning one by one to generate the measurement matrix $\mathbf{A}$. The initialization of $\mathbf{o}$ is (Fig. \ref{fig:result8}a1) obtained by raster-scanning the object without correcting for the scattering layer. In each iteration, we solve Eq.\eqref{eq:eq3} first, from which we retrieve $\mathbf{T}^{'}$, itself containing the wavefront $\mathbf{t}$. Then, $\rm{arg}(\mathbf{t}^{*})$ is displayed on the SLM to correct the scattering layer for raster-scanning to obtain an update of $\mathbf{o}$ (Fig. \ref{fig:result8}a2). This process repeats until a desired stop criteria is achieved (in our case, a strong signal). Typically 3 iterations are needed to converge to a single focus as shown by the inspection camera at each iteration (Fig.\ref{fig:result8}b1-3). To ensure fast convergence, the dimension of $\mathbf{o}$ is set equal to $N_{\rm{slm}}$. Comparison of a high-resolution imaging taken after the optimization procedure (Fig. \ref{fig:result8}c2) shows excellent agreement with the inspection camera (Fig. \ref{fig:result8}c3).

\begin{figure}[ht]
    \centering
	\includegraphics[width=0.7\columnwidth]{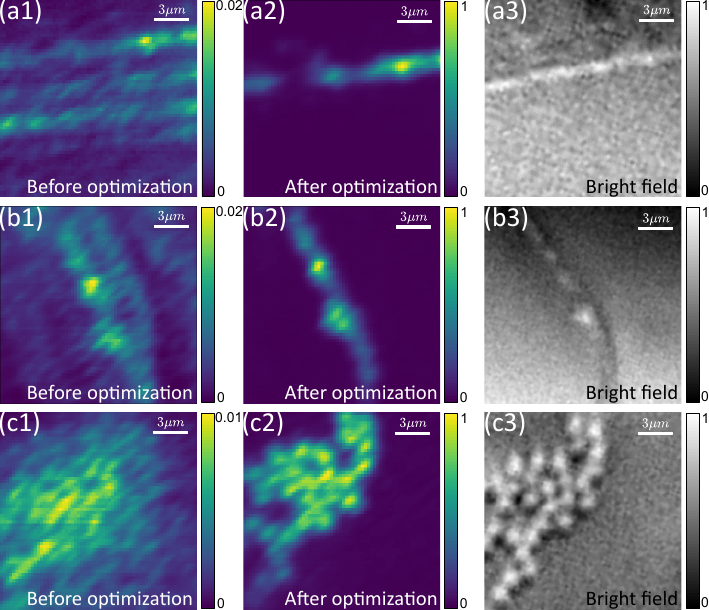}
	\caption{ \textbf{Retrieval of $\mathbf{T}$ for extended objects.} The samples in (a) and (b) are fibres that coming from lens tissue marked by a fluorescence pen. The samples in (c) is a bunch of 1$\mu$m beads. (a1), (b1) and (c1) are the imaging result by scanning in $64\times64$ steps without wavefront correction. (a2), (b2) and (c2) are the corresponding results after the wavefront correction. Here, we use 16 random patterns scanning to generate the measurement matrix $\mathbf{A}$, and put $\mathbf{A}$ and $\mathbf{y}$ as input for Alg.\ref{alg:alg2}. The samples in (a), (b) take 3 iterations to converge, and (c) takes 5 iterations. (a3), (b3) and (c3) are the corresponding bright field images.}
	\label{fig:result16}
\end{figure}

Inspection at each iteration shows that the speckle converges to a single foci despite the size of the object. Fig.\ref{fig:result16} shows the experimental results for extended objects. Fig.\ref{fig:result16}a-b are fluorescence-dyed fibres into orthogonal directions spanning the whole speckle envelope. Before optimization, clear replica of the object are seen because of the sparse 2PF speckle excitation. After optimization, convergence to a single focus is ensured. Even in extreme cases where the sample extend quasi homogeneously the speckle envelope (Fig.\ref{fig:result16}c), there is clearly a convergence to a single focus as shown by comparison with brightfield images (Fig.\ref{fig:result16} c3, control camera). In these examples, because we reconstruct an extended complex $\mathbf{o}$, to speed up convergence, we use 16 random patterns raster-scanning to generate $\mathbf{A}$ which weakly scales with the object complexity. The three results showing here take 3, 3 and 5 iterations to converge, respectively.

\section*{Discussion}
We presented a method to non-invasively extract the transmission matrix of a buried fluorescent object, using a single-pixel detector. To validate the principle, we conducted an experiment by placing a single-pixel detector on one side of the object. However, in practice, it can also be placed on the other side to enable epi-detection mode. In both scenarios, the forward model of the system remains the same. Based on a phase-retrieval formalism, we exploit two extreme cases on the level of correlations embedded in $\mathbf{T}$. If no correlations exists (as it would be the case for a very optically-thick medium), we show that retrieval of $\mathbf{T}$ is unique for nonlinear optical processes, such as two-photon fluorescence signals. 
The downside of this approach is that it requires a number of measurements that scales with the object complexity, which leads to lengthy acquisitions. We can largely circumvent this limitation if $\mathbf{T}$ displays correlations, in the form of memory effect, by introducing a tilt matrix $\mathbf{F}^{T}$. Rigorous solution is achieved by updating $\mathbf{o}$ at each new estimation of $\mathbf{T}$. In this case, the number of wavefront measurements does not scale with object sparsity, so measurements can be significantly faster.

Two kind of methods exist for two-photon imaging through scattering medium. The first is known as wavefront shaping based method \cite{tang2012superpenetration,bridges1974coherent}, which need fast SLMs in order to ultimately outpace the speckle decorrelation time imposed by living tissue. The second is focus scanning holographic aberration probing (F-SHARP) \cite{papadopoulos2017scattering} and can be viewed as an interferometric method. 
Both, and works that follows \cite{may2021fast,papadopoulos2020dynamic}, are optimization based methods and can only image within the memory effect range. We propose a method for imaging beyond the memory effect range, although it needs more measurements than optimization based methods, the data processing is not needed between the measurements which leaves a great possibility to improve the measurement speed. In addition, due to the random illumination strategy, the proposed method can prevent bleach compared to optimization based methods. A particular advantage here is that, when the imaging object is smaller than the ME range, the proposed method can realize faster focusing speed compared to wavefront shaping methods as in principle galvanometric and digital micromirror devices can be used, and have better stability compared to interferometric methods.

\section*{Methods}
\subsection*{Experimental setup and sample preparation}
The experimental setup is sketched in Fig.\ref{fig:setup}. The source used is an OPO (Coherent Mira OPO-X) pumped by a pulsed Ti-sapphire laser (Coherent Chameleon Ultra II) with 140 fs pulse duration and repetition rate of 80 MHz. For 2P excitation, the OPO is tuned to 1050 nm. To modulate the wavefront, we use a liquid crystal phase-only SLM (Meadowlarks HSP512L) which is imaged on the back focal plane of a water immersion objective (Zeiss W Plan-Apochromat 40x/1.0 DIC M27). After illuminating the sample, a second microscope objective (Zeiss EC Plan-NEOFLUAR 40x/1.3) images the excitation light pattern onto a CCD camera (Basler acA1300-30$\rm{\mu}$m) and collects the fluorescence light which is steered towards a PMT (Hamamatsu H7422P-40) by a dichrocic mirror (Semrock Di03-R785-t1-25x36). Additional filters in front of the PMT (Thorlabs FESH0600, FESH0700) ensure that only nonlinear fluorescent light is detected.

For the scattering medium, a rough surface is introduced on a \#1.5 cover slip by sandblasting its bottom surface with 220 grit sand. The beads (Fluospheres carboxylate, 1.0$\rm{\mu}$m, red (580nm/605nm)) are drop-casted on the top surface of the coverslip, about 170$\rm{\mu}$m above the scattering layer. They are covered by the NOA65 glue to create a near-index matched environment. Then, a second \#1.5 cover slip is placed on top of the beads. Finally, the glue is cured under UV light, the curing process is kept short to avoid bleaching the beads. The fibres are provided by Thorlab lens tissue marked by a orange fluorescence pen, they are directly covered by a second \#1.5 cover slip on top.  

\subsection*{Reconstruction details}

\begin{algorithm}[H]
\DontPrintSemicolon
\SetAlgoLined
\SetKwInOut{Input}{input}\SetKwInOut{Output}{output}
\Input{number of gradient descent, $nn$; step size, $\Delta$; measurement matrix, $\mathbf{A}$;  measurement vector, $\mathbf{y}$; column number, $N^{'}_{\rm{obj}}$.}
\Output{columns of transmission matrix, $\mathbf{t}_{i}$ ($i = 1,2\cdots N^{'}_{\rm{obj}}$).}
\For{count $\leftarrow 1$ \KwTo $nn$}{

    $\mathbf{t}_{i} = \mathbf{t}_{i} - \Delta \frac{\frac{\partial f}{\partial \mathbf{t}^{*}_{i}} } {\lVert \frac{\partial f}{\partial \mathbf{t}^{*}_{i}} \rVert}_{2}  $ \;
}
\caption{Multiplexed phase retrieval}
\label{alg:alg1}
\end{algorithm}
In order to reconstruct $\mathbf{t}_{i}$ in Eq.\eqref{eq:eq2} when $p=2$. The Alg.\ref{alg:alg1} is proposed to solve the following optimized problem:
\begin{equation}
\rm{min}_{\mathbf{t}_{i}} f,\quad \rm{with} \quad f =  \lVert \mathbf{y} - \sum_{i=1}^{N^{'}_{obj}}\lvert \mathbf{A}\mathbf{t}_{i} \rvert^{4} \rVert^{2}_{2},
\label{eq:eq4}
\end{equation}
 where $N^{'}_{\rm{obj}}$ is chosen to be equal or larger than the actual number of sources behind the scattering medium. Some of the columns of the reconstructed $\mathbf{T}$ will be repeated if $N^{'}_{\rm{obj}}$ is chosen larger (see Supplementary Information for more details). For the initialization, we randomly initialize each column $\mathbf{t}_{i}$ of the transmission matrix. According to \cite{remmert1991theory,fischer2005precoding,bian2015fourier}, the derivative of the complex quadratic cost function with
respect to $\mathbf{t}^{*}_{i}$, i.e. $\frac{\partial f}{\partial \mathbf{t}^{*}_{i}}$, is necessary for updating $\mathbf{t}_{i}$ in each iteration, and $\frac{\partial f}{\partial \mathbf{t}^{*}_{i}}$ can be easily caculated according to the Wirtinger derivatives as
\begin{equation}
\frac{\partial f}{\partial \mathbf{t}^{*}_{i}} = 4\mathbf{A}^{\dag}[(\sum_{i=1}^{N^{'}_{\rm{obj}}} \lvert \mathbf{A}\mathbf{t}_{i} \rvert^{4}-\mathbf{y}) \odot {\lvert \mathbf{A}\mathbf{t}_{i} \rvert}^{2} \odot (\mathbf{A}\mathbf{t}_{i})].
\label{eq_grad1}
\end{equation}
 For the result in Fig.\ref{fig:fig2}, $16\times N^{'}_{\rm{obj}} \times N_{\rm{slm}}$ random patterns are displayed on the SLM to obtain $\mathbf{y}$, where $N_{\rm{slm}} = 16\times16$ and $N^{'}_{\rm{obj}} =7$. The reconstruction of Alg.\ref{alg:alg1} takes less than 10 seconds on GPU.

\begin{algorithm}[H]
\DontPrintSemicolon
\SetAlgoLined
\SetKwInOut{Input}{input}\SetKwInOut{Output}{output}
\Input{number of gradient descent, $nn$; step size, $\Delta$; measurement matrix, $\mathbf{A}$;  measurement vector, $\mathbf{y}$.}
\Output{aberration, $\mathbf{t}$; object, $\mathbf{o}$.}
\While{2-photon signal is low}{
\For{count $\leftarrow 1$ \KwTo $nn$}{

    $\mathbf{t} = \mathbf{t} - \Delta \frac{\frac{\partial f}{\partial \mathbf{t}^{*}} } {\lVert \frac{\partial f}{\partial \mathbf{t}^{*}} \rVert_{2}}  $ \;
}

Update $\mathbf{o}$ by raster-scanning the sample with the compensate wavefront $\rm{arg}(\mathbf{t}^{*})$ 

}
\caption{Phase retrieval + Scanning}
\label{alg:alg2}
\end{algorithm}

In order to reconstruct $\mathbf{T}^{'}$ concatenated by vector $\mathbf{t}$ in Eq.\eqref{eq:eq3}, the Alg.\ref{alg:alg2} is proposed to solve the following optimized problem:
\begin{equation}
\rm{min}_{\mathbf{t},\mathbf{o}} f,\quad \rm{with}\quad f = \sum^{N_{m}}_{q=1} ( y_{q} - \mathbf{o}^{T}\lvert \mathbf{F}(\mathbf{a}^{*}_{q} \odot \mathbf{t}) \rvert^{4} )^{2},
\label{eq:eq5}
\end{equation}
where $\mathbf{a}^{\dag}_{q}$ represents the $q$th row of $\mathbf{A}$, $y_q$ represents the $q$th element of $\mathbf{y}$. For the initialization, $\mathbf{t}$ is randomly initialized, and $\mathbf{o}$ is obtained by raster-scanning the object with a scanning device (applying phase ramps on the SLM in our case). The gradient $\frac{\partial f}{\partial \mathbf{t}^{*}}$ is derived as: 
\begin{equation}
\begin{split}
\frac{\partial f}{\partial \mathbf{t}^{*}} &= 4\sum^{N_{m}}_{q=1} (\mathbf{o}^{T}\lvert \mathbf{F}(\mathbf{a}^{*}_{q} \odot \mathbf{t}) \rvert^{4} - y_{q})\mathbf{a}_{q}\\
&\odot \{ \mathbf{F}^{\dag}[\mathbf{o}\odot \lvert \mathbf{F}(\mathbf{a}^{*}_{q} \odot t) \rvert^{2} \odot (\mathbf{F}(\mathbf{a}^{*}_{q} \odot \mathbf{t})) ] \}.
\label{eq_grad2}
\end{split}
\end{equation}
However, the auto-gradient derived by Pytorch are used in our code for faster computation speed. In each iteration, $nn$ times of gradient descent are performed to reconstruct the $\mathbf{t}$, then the $\mathbf{o}$ is updated by raster-scanning the sample with a scanning device and displaying compensate pattern $\rm{arg}(\mathbf{t}^{*})$ on the SLM simultaneously (applying a complete set of phase ramps on top of $\rm{arg}(\mathbf{t}^{*})$ on the SLM in our case). For the results in Fig.\ref{fig:result8} and Fig.\ref{fig:result16}, $8 \times N_{\rm{slm}}$ and $16 \times N_{\rm{slm}}$ measurements are taken to get $\mathbf{y}$ respectively. They are obtained by using 8 and 16 random patterns on the SLM and raster-scanning the sample with them one by one, $N_{\rm{slm}} = 16\times16$ measurements for each raster-scanning. The sampling strategy here can significantly reduce the sampling time if the galvo-mirror are used as scanning device and placed on the conjugate plane of the SLM. The higher resolution images in Fig.\ref{fig:result8} and Fig.\ref{fig:result16} are obtained by raster-scanning the sample with SLM in $64\times64$ steps. In each iteration, the reconstruction of $\mathbf{t}$ take less than 10 seconds on GPU. 

The computational resource we used: GPU, 2 Nvidia GeForce RTX 2080 Ti; RAM, 64 GB.

\section*{Code available}
Simulation codes are available at https://github.com/comediaLKB/Single\_pixel\_TM\_recovery.

\section*{Acknowledgements}
L.V. acknowledges funding from the Swiss National Science Foundation (project P400P2\_199329). S.Z. acknowledges funding from the China Scholarship Council (CSC) (202006280163). This project was funded by the European Union’s Horizon 2020 research and innovation program under the FET-Open grant No. 863203 (Dynamic). The authors acknowledge the assistance of Eric Bezzam in optimizing the code.

\section*{Competing interests}
The authors declare no competing interests.

\bibliography{scifile}

\bibliographystyle{Science}

\clearpage

\section*{Supplementary Information:}

\subsection*{The uniqueness of the solution depending on $p$}
To aid in understanding, we consider a scenario that only 2 beads under the scattering medium. We aim to demonstrate that, given a sufficient number of random measurements, the solution to Eq.\eqref{eq:eq2} in the main text is non-unique when $p=1$, while it is unique when $p=2$. It‘s not hard to see that the conclusion can be generalized to the case of more than 2 beads. 

The corresponding 2 columns of $\mathbf{T} \in \mathbb{C}^{N_{\rm{slm}}\times2}$ are $\mathbf{t}_{1}$ and $\mathbf{t}_{2}$, respectively. $\mathbf{a}^{\dag} \in \mathbb{C}^{1\times N_{\rm{slm}}}$ represents one row of the measurement matrix $\mathbf{A}$. and the corresponding fluorescence signal $y$ is: 
\begin{equation}
y = \lvert \mathbf{a}^{\dag}\mathbf{t}_{1} \rvert^{2p} + \lvert \mathbf{a}^{\dag}\mathbf{t}_{2} \rvert^{2p},
\label{eq:prov1}
\end{equation}
Assuming 
$\mathbf{t}^{\dag}_{1}\mathbf{t}_{2} = 0$, which means the columns
of transmission matrix are orthogonal to each other (since the transmission matrix are complex random matrix, this a weak assumption). We can then represent inner product in a canonical basis starting with 
\begin{equation}
 \mathbf{l}_{1} =
 \begin{pmatrix}
 1\\
 0\\
 0\\
 \vdots
 \end{pmatrix}
\quad \rm{and} \quad
 \mathbf{l}_{2} =
 \begin{pmatrix}
 0\\
 1\\
 0\\
 \vdots
 \end{pmatrix}.
\label{eq:prov2}
\end{equation}
Eq.\eqref{eq:prov1} in this basis becomes:
\begin{equation}
\begin{split}
y &= \lvert (\mathbf{R}\mathbf{a})^{\dag}\mathbf{R}\mathbf{t}_{1} \rvert^{2p} + \lvert (\mathbf{R}\mathbf{a})^{\dag}\mathbf{R}\mathbf{t}_{2} \rvert^{2p}\\
  &= \lVert\mathbf{t}_{1}\rVert^{2p} \lvert \mathbf{\hat{a}}^{\dag}\mathbf{l}_{1}  \rvert^{2p} + 
     \lVert\mathbf{t}_{2}\rVert^{2p} \lvert \mathbf{\hat{a}}^{\dag}\mathbf{l}_{2} \rvert^{2p}\\
  &= \lVert\mathbf{t}_{1}\rVert^{2p} \lvert \hat{a}_{1} \rvert^{2p} + \lVert\mathbf{t}_{2}\rVert^{2p} \lvert \hat{a}_{2} \rvert^{2p},
\end{split}
\label{eq:prov3}
\end{equation}
where $\mathbf{R}$ represents a unitary basis rotation matrix, and $\hat{a}_{i}$ represents the $i$th element of the vector $\mathbf{\hat{a}}$.

The question now is: Are there any $(\mathbf{u}_{1},\mathbf{u}_{2})\neq(\mathbf{l}_{1},\mathbf{l}_{2}) $ that satisfy the following formula?
\begin{equation}
\forall \mathbf{\hat{a}}^{\dag} \quad
\lVert\mathbf{t}_{1}\rVert^{2p}\lvert \hat{a}_{1} \rvert^{2p} + \lVert\mathbf{t}_{2}\rVert^{2p}\lvert \hat{a}_{2} \rvert^{2p} = \lvert \mathbf{\hat{a}}^{\dag}\mathbf{u}_{1}  \rvert^{2p} + \lvert \mathbf{\hat{a}}^{\dag}\mathbf{u}_{2} \rvert^{2p}.
\label{eq:prov4}   
\end{equation}

\textbf{In the $p=1$ case}, for any $2\times2$ unitary matrix $\mathbf{S}$ we can find a set of solution:
\begin{equation}
\begin{pmatrix}
\mathbf{u}_{1}\\
\mathbf{u}_{2}\\
\end{pmatrix}  
= \mathbf{S}
\begin{pmatrix}
\lVert\mathbf{t}_{1}\rVert\mathbf{l}_{1}\\
\lVert\mathbf{t}_{2}\rVert\mathbf{l}_{2}\\
\end{pmatrix}  
\label{eq:prov5}  
\end{equation}
satisfy Eq.\eqref{eq:prov4}. For example, $(\mathbf{u}_{1},\mathbf{u}_{2}) = (\frac{1}{\sqrt{2}}(\lVert\mathbf{t}_{1}\rVert\mathbf{l}_{1} + \lVert\mathbf{t}_{2}\rVert\mathbf{l}_{2}), \frac{1}{\sqrt{2}}(\lVert\mathbf{t}_{1}\rVert\mathbf{l}_{1} - \lVert\mathbf{t}_{2}\rVert\mathbf{l}_{2}))$. Thus, the Eq.\eqref{eq:eq2} in the main text has infinite solution when $p=1$.

\textbf{In the $p=2$ case}, Eq.\eqref{eq:prov4} becomes
\begin{equation}
\forall \mathbf{\hat{a}}^{\dag} \quad
\lVert\mathbf{t}_{1}\rVert^{4} \lvert \hat{a}_{1} \rvert^{4} + \lVert\mathbf{t}_{2}\rVert^{4}\lvert \hat{a}_{2} \rvert^{4} = \lvert \mathbf{\hat{a}}^{\dag}\mathbf{u}_{1}  \rvert^{4} + \lvert \mathbf{\hat{a}}^{\dag}\mathbf{u}_{2} \rvert^{4}.
\label{eq:prov6}   
\end{equation}
Let $\hat{a}_{1} = \hat{a}_{2} = 0$ and $\hat{a}_{i} = \forall, i\neq 1\mkern8mu\rm{or}\mkern8mu 2$, we have
\begin{equation}
\mathbf{u}_{1} = 
\begin{pmatrix}
\alpha\\
\beta\\
0\\
0\\
\vdots\\
\end{pmatrix}  
\quad \mathbf{u}_{2} = 
\begin{pmatrix}
\gamma\\
\delta\\
0\\
0\\
\vdots\\
\end{pmatrix}. 
\label{eq:prov7}  
\end{equation}
Let $\hat{a}_{1} = \forall$, $\hat{a}_{2} = 0$, we have
\begin{equation}
\lvert \alpha \rvert^{4} + \lvert \gamma  \rvert^{4} = \lVert\mathbf{t}_{1}\rVert^{4}.
\label{eq:prov8}    
\end{equation}
Let $\hat{a}_{1} = 0$, $\hat{a}_{2} = \forall$, we have
\begin{equation}
\lvert \beta \rvert^{4} + \lvert \delta  \rvert^{4} = \lVert\mathbf{t}_{2}\rVert^{4}.
\label{eq:prov9}    
\end{equation}
Let $\hat{a}_{1} = \hat{a}_{2} = \forall$, we have
\begin{equation}
\lvert \alpha + \beta \rvert^{4} + \lvert \gamma + \delta  \rvert^{4} = \lVert\mathbf{t}_{1}\rVert^{4}+\lVert\mathbf{t}_{2}\rVert^{4}.
\label{eq:prov10}    
\end{equation}
Let $\hat{a}_{1} = -\hat{a}_{2} = \forall$, we have
\begin{equation}
\lvert \alpha - \beta \rvert^{4} + \lvert \gamma - \delta  \rvert^{4} = \lVert\mathbf{t}_{1}\rVert^{4}+\lVert\mathbf{t}_{2}\rVert^{4}.
\label{eq:prov11}    
\end{equation}
Eq.\eqref{eq:prov10} plus Eq.\eqref{eq:prov11} and using Eq.\eqref{eq:prov8} and Eq.\eqref{eq:prov9} we have
\begin{equation}
(\alpha\beta^{*})^2 + (\alpha^{*}\beta)^2 + 4\lvert \alpha^{*}\beta \rvert^{2} + (\gamma\delta^{*})^2 + (\gamma^{*}\delta)^2 + 4\lvert \gamma^{*}\delta \rvert^{2} = 0.
\label{eq:prov12}    
\end{equation}
Defining 
\begin{equation}
\alpha\beta^{*} = m + in, \quad \alpha^{*}\beta = m - in,\quad \gamma\delta^{*} = p + iq, \quad \gamma^{*}\delta = p - iq,
\label{eq:prov13}    
\end{equation}
into Eq.\eqref{eq:prov12}, where $i$ represents the imaginary unit and $m,n,p,q$ are real, we have
\begin{equation}
6m^{2} + 2n^{2} + 6p^{2} + 2q^{2} = 0. 
\label{eq:prov14}    
\end{equation}
Thus, $m=n=p=q=0$. It means $\alpha\beta^{*} =  \gamma\delta^{*} = 0$. So $\alpha \mkern8mu\rm{or}\mkern8mu \beta =0$ and $\gamma \mkern8mu\rm{or}\mkern8mu \delta =0$. Therefore, we conclude that  
\begin{equation}
(\mathbf{u}_{1},\mathbf{u}_{2}) = (\mathbf{l}_{1},\mathbf{l}_{2})\mkern8mu\rm{or}\mkern8mu(\mathbf{l}_{2},\mathbf{l}_{1}).
\label{eq:prov15}    
\end{equation}
Thus, Eq.\eqref{eq:eq2} in the main text has unique solution when $p=2$. 

\subsection*{$\mathbf{T}$ reconstruction quality for using different $N^{'}_{\rm{obj}}$}

\begin{figure}[htp]
        \centering
	\includegraphics[width=0.9\columnwidth]{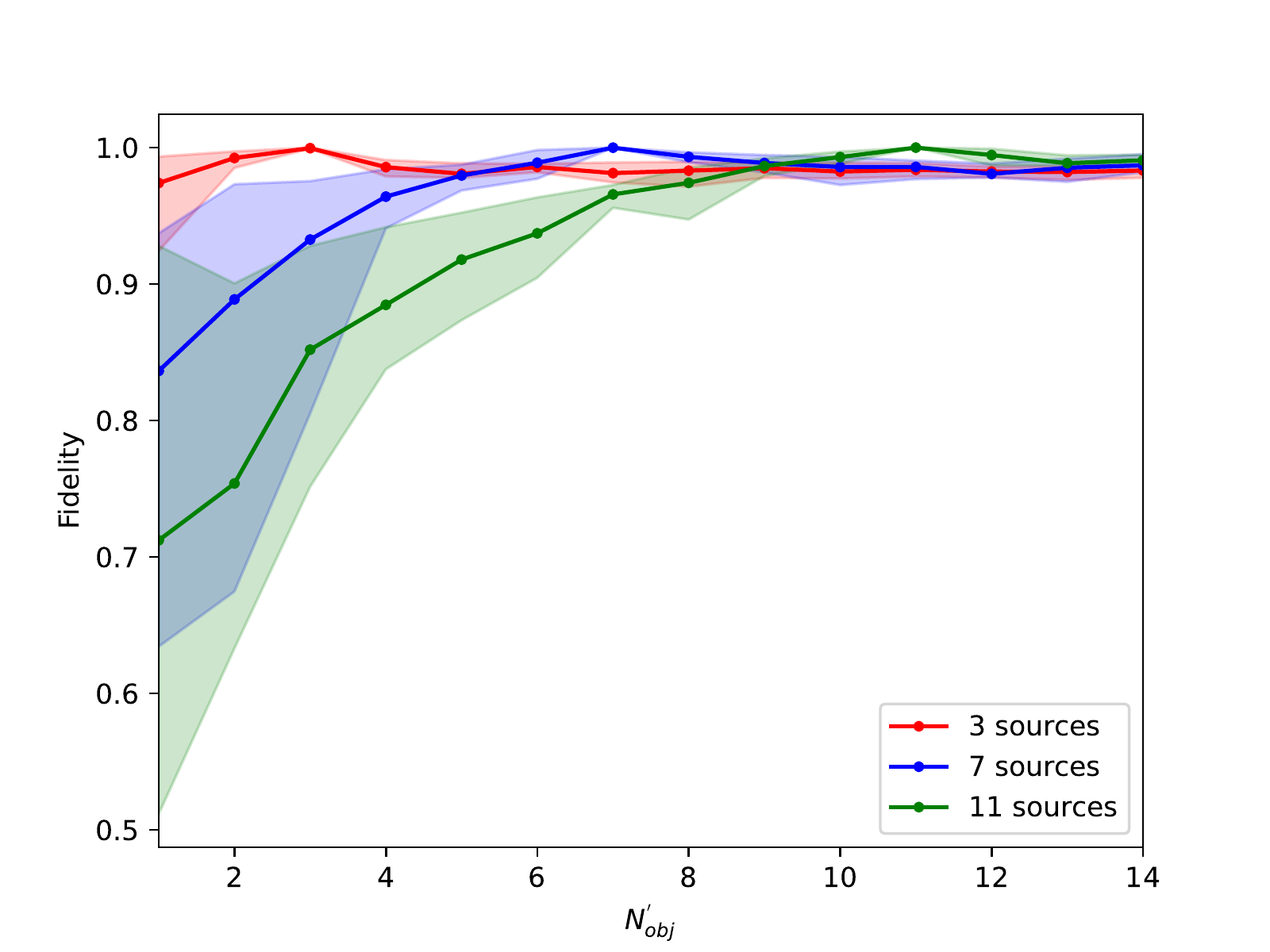}
	\caption{\textbf{The fidelity of retrieval $\mathbf{T}$ with different $N^{'}_{\rm{obj}}$}: Three different kind of random matrix $\mathbf{T}$ are generated for ground truth, they have 3 (red data), 7 (blue data) and 11 (green data) columns respectively. $\mathbf{T}$ reconstruction fidelity are evaluated when using different $N^{'}_{\rm{obj}}$. The each solid line represents the average of 10 realization, and the shadow represents the fidelity range distribution.}
	\label{fig:fidelity}
\end{figure}

In order to estimate the $\mathbf{T}$ reconstruction quality using different $N^{'}_{\rm{obj}}$, three kind of random matrix $T$ are generated for ground truth, the number of columns they contain is  3, 7 and 11 respectively. Different $N^{'}_{\rm{obj}}$ ranging from 1 to 14 are using in Alg.\ref{alg:alg1} to reconstruct $\mathbf{T}$. The reconstruction fidelity is shown in Fig.\ref{fig:fidelity}, the red, blue and green are corresponding to the ground truth $\mathbf{T}$ with 3, 7 and 11 columns respectively. 

The fidelity of each column $\mathbf{t}_i$ is calculated between the reconstructed $\mathbf{t}_i$ and every column of the ground-truth $\mathbf{T}$ then pick the maximum. The fidelity of the transmission matrix $\mathbf{T}$ is the average fidelity of all the column $\mathbf{t}_i,i=1,2,\cdots,N^{'}_{\rm{obj}}$. As showing in the result, when the $N^{'}_{\rm{obj}}$ is chosen smaller than the sources number, the reconstructed column $\mathbf{t}_i$ is not guaranteed to match one of the column in the ground-truth matrix $\mathbf{T}$. On the contrary, when the $N^{'}_{\rm{obj}}$ is chosen bigger than the sources number, each reconstructed $t_i$ match one of the column in the ground-truth $\mathbf{T}$. Actually, in this case some of the columns of the reconstructed $\mathbf{T}$ will be repeated. In the simulation, $N_{m} = 65536$ and $N_{\rm{slm}} =16\times16$. The each solid line represents the average of 10 realization, and the shadow represents the fidelity range distribution.

\subsection*{Transmission matrix based image reconstruction}

\begin{figure}[htp]
        \centering
	\includegraphics[width=\columnwidth]{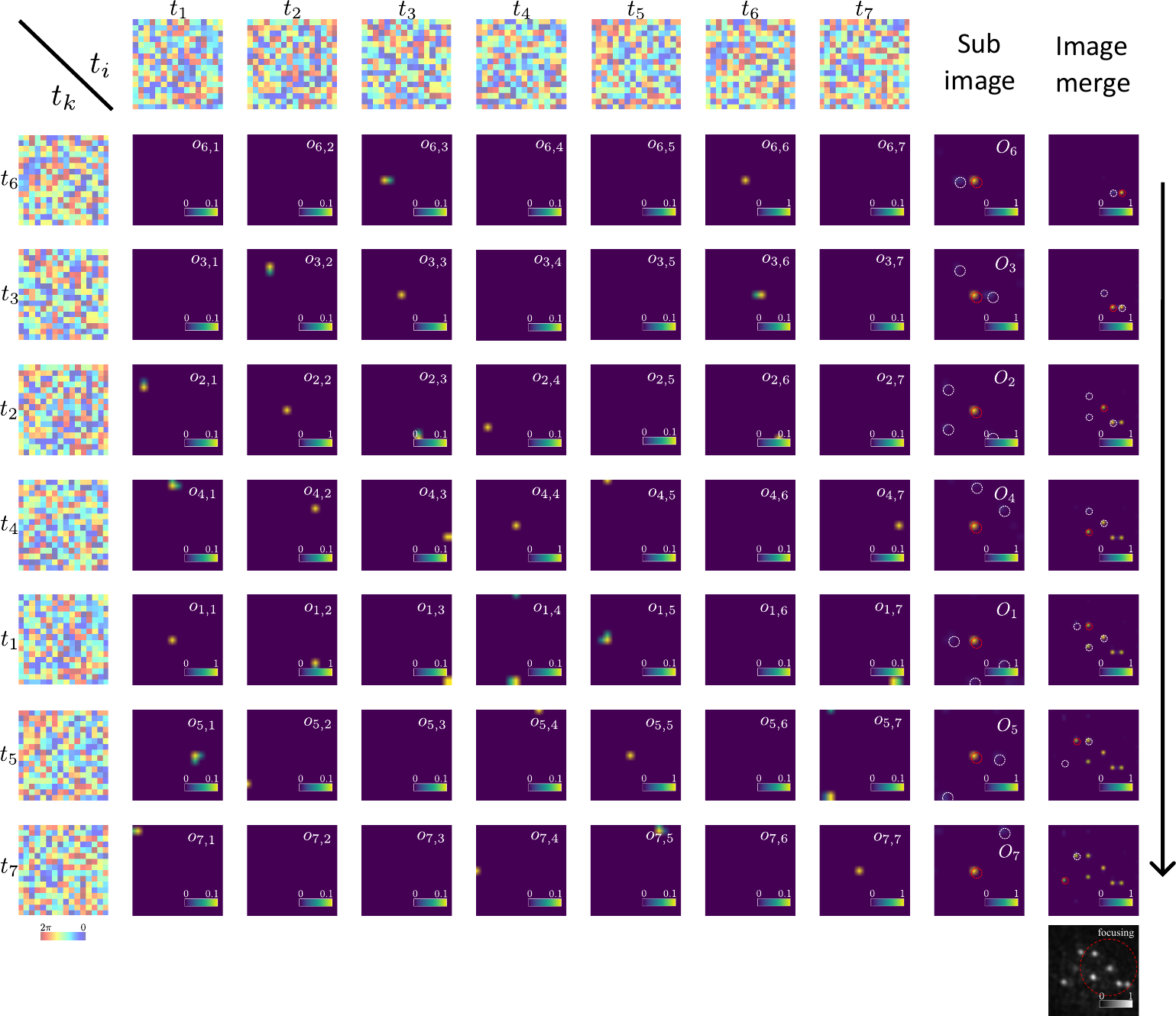}
	\caption{\textbf{Imaging procedure of Fig.\ref{fig:fig2}e in the manuscript}: At first, for each column $\mathbf{t}_{i}$, the relative position map $o_{i,k}$ with column $\mathbf{t}_{k}$ is calculated and thresholded. Then, the sub images $O_{i}$ centered by the emitter corresponding to $\mathbf{t}_{i}$ are obtained by summing all the corresponding relative position maps. Finally, the global image is merged by all the sub images.}
	\label{fig:Imaging}
\end{figure}

Previous works that realized imaging beyond the memory effect (ME) by analysing the retrieved distance map between all the emitters \cite{boniface2020non,zhu2022large,d2022physics,soldevila2023functional}, but they all used a camera to retrieve all the emission fluorescent patterns first, which is not sensitive enough for non-linear microscopy. Here, we retrieved the relative position between every two emitters by using the transmission matrix $\mathbf{T}$. Fig.\ref{fig:Imaging} shows the image reconstruction work flow by analysing the $T$ we retrieved in result Fig.\ref{fig:fig2}e in the manuscript. If the two emitters are really close, the corresponding two columns in $\mathbf{T}$ should be highly correlated and provide relative position information. The relative position map of the two emitters is calculated by:

\begin{equation}
o_{i,k} = \lvert \mathcal{F}_{2d}( \frac{\mathbf{t}^{*}_{i}}{\lvert \mathbf{t}^{*}_{i} \rvert }  \odot \frac{\mathbf{t}_{k}}{\lvert \mathbf{t}_{k} \rvert}  )\rvert^2,
\label{eq:eqs3}
\end{equation}
where $\mathcal{F}_{2d}$ represent the 2d Fourier transform, $\odot$ means element wise production. To infer whether the columns $\mathbf{t}_i$ and $\mathbf{t}_k$ are within or beyond the ME range, we study $\alpha = \frac{max \{ o_{i,k} \}}{max \{ o_{k,k} \}}$ as a function of relative distance, where $max \{ o_{i,k} \}$ stands for maximum value of $o_{i,k}$. A given threshold, $\alpha_{tres}$ of $\alpha$ is introduced to evaluate it. For example, if $\alpha$ is greater than $\alpha_{tres}$, the two columns belong to the same ME range. Otherwise, they belong to different ME ranges, the relative position map $o_{i,k}$ is reassigned to a zero matrix. The pairwise relative position map $o_{i,k}$ after thresholding is showing in Fig.\ref{fig:Imaging}.

Then the sub images of the object in the vicinity of the emitter $k$ are obtained by $O_{k} = \sum_{i=1}^{N^{'}_{\rm{obj}}} o_{i,k}$. In order to merge the sub images to a global image, we also need the relative position vector $\mathbf{r}_{i,k}$ of the two emitters $i$ and $k$. If emitters $i$ and $k$ are in the same ME range, $\mathbf{r}_{i,k}$ is a vector in the map $o_{i,k}$ that origin at the center and end at the narrow delta-like peak. If emitters $i$ and $k$ are beyond the ME range but emitter $j$ is between them, we can still calculate the $\mathbf{r}_{i,k}$ between them as $\mathbf{r}_{i,k} = \mathbf{r}_{i,j} + \mathbf{r}_{j,k}$. Therefore, even if the ensemble of emitters expands well beyond the ME range, the full spatial distribution can be recovered if the different isoplanatic patches are “connected" by emitters.

\begin{figure}[htp]
        \centering
	\includegraphics[width = 0.7\columnwidth]{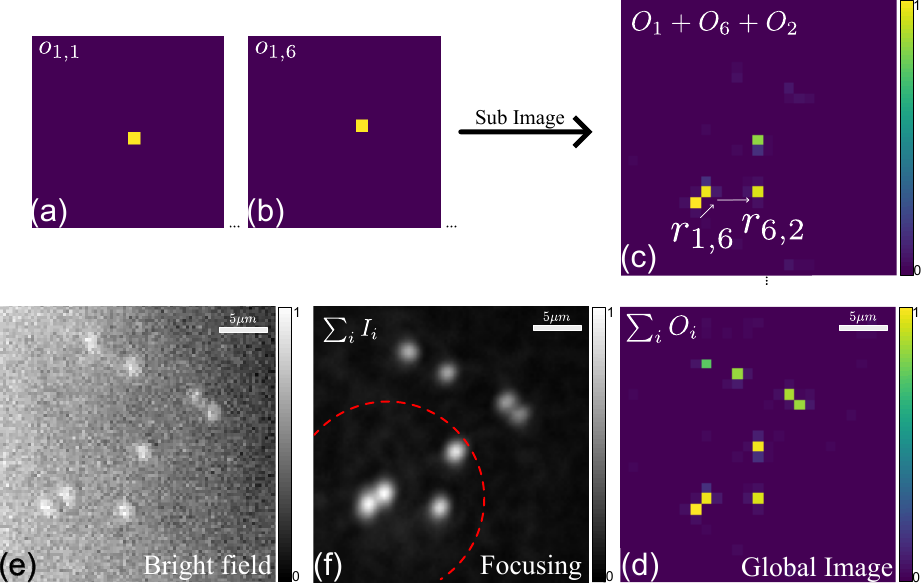}
	\caption{\textbf{Image merge strategy based on relative position vector $\mathbf{r}_{i,k}$.} (a) and (b) are relative position maps $o_{1,1}$ and $o_{1,6}$. (c) 3 sub images $O_{1},O_{6},O_{2}$ are stitched by 2 relative position vector $\mathbf{r}_{1,6}$ and $\mathbf{r}_{6,2}$. (d) The $O_{global}$ after image merge. (e) Brightfield image taken with control camera confirms the number of sources. (f) Sum of all images for each focus demonstrating unique single focus for each column retrieved.}
	\label{fig_SP_TM_image_merge2}
\end{figure}

\begin{figure}[htp]
        \centering
	\includegraphics[width = 0.7\columnwidth]{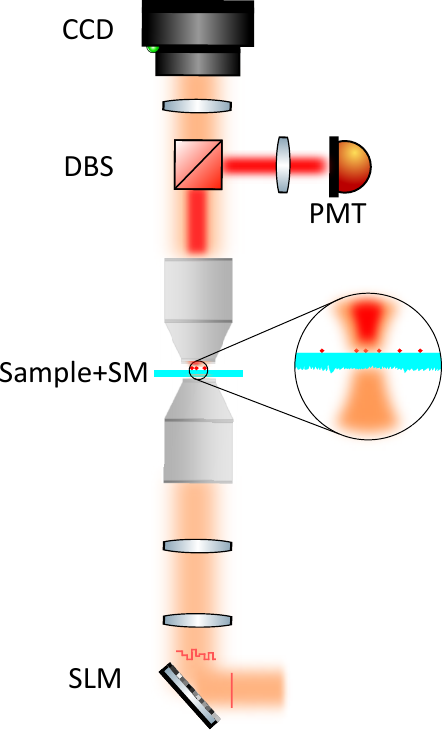}
	\caption{\textbf{Experimental setup}: SLM, spatial light modulator; Sample, sample; SM, scattering medium; DBS, dichroic beam splitters; CCD, charge-coupled device; PMT, photo-multiplier tube}
	\label{fig:setup}
\end{figure}

Finally, we use the method proposed in \cite{zhu2022large} to merge all the sub images, and realize imaging beyond ME range. In short, the global reconstruction of $O_{global}$ can be obtained by composing all the sub images $O_{k}$ into one images, taking into account their relative positions with respect to the first emitter, ${\mathbf{r}}_{k,1}$:
\begin{equation}
O_{global} = \sum_{k=1}^{N^{'}_{\rm{obj}}}O_{k}(\mathbf{r}-\mathbf{r}_{k,1}).
\label{eq:eqs4}
\end{equation}
Therefore, a image beyond the ME range is obtained. The red dashed circle showing in the reference focusing image in Fig.\ref{fig:Imaging} represents the ME range. Another result with 8 fluorescence beads is shown in Fig.\ref{fig_SP_TM_image_merge2}.

\clearpage

\end{document}